\documentstyle[12pt]{article}

\def\thefootnote{\fnsymbol{footnote}}
\def\11{\mbox{$1$}}
\renewcommand{\thefootnote}{\alph{footnote}}

\newcommand{\rref}[1]{(\ref{#1})}
\newcommand{\beqn}{\begin{equation}}
\newcommand{\eeqn}{\end{equation}}
\newcommand{\beqarr}{\begin{eqnarray}}
\newcommand{\eeqarr}{\end{eqnarray}}

\newcommand{\matc}{\begin{array}{c}}
\newcommand{\matcc}{\begin{array}{cc}}
\newcommand{\matccc}{\begin{array}{ccc}}
\newcommand{\matcccc}{\begin{array}{cccc}}
\newcommand{\emat}{\end{array}}

\newcommand{\Lag}{L}

\newcommand{\IH}{\relax{\rm I\kern-.18em H}}
\newcommand{\IR}{\relax{\rm I\kern-.18em R}}
\newcommand{\IK}{\relax{\rm I\kern-.18em K}}
\newcommand{\II}{\hbox{\rm 1\kern-.35em 1}}
\newcommand{\Is}{\relax{\rm 1\kern-.35em 1}}

\begin{document}

\begin{titlepage}

April 2000         \hfill
%\vskip -0.55cm 
%\hfill    UCB-PTH-00/xx  
%\\
\vskip -0.25cm 
\hfill  LBNL-45360    
\begin{center}
\vskip .15in
\renewcommand{\thefootnote}{\fnsymbol{footnote}}
{\large \bf Proof of a Symmetrized Trace Conjecture for the Abelian 
Born-Infeld Lagrangian}
\vskip .25in
Paolo Aschieri\footnote{present
address: Sektion Physik LMU, Theresienstr. 37, D-80333 
M\"{u}nchen}\footnote{email 
address: aschieri@theorie.physik.uni-muenchen.de},
Daniel Brace\footnote{email address: DMBrace@lbl.gov}, 
Bogdan Morariu\footnote{email address: BMorariu@lbl.gov} and 
Bruno Zumino\footnote{email address: zumino@thsrv.lbl.gov}
\vskip .25in

{\em    Department of Physics  \\
        University of California   \\
                                and     \\
        Theoretical Physics Group   \\
        Lawrence Berkeley National Laboratory  \\
        University of California   \\
        Berkeley, California 94720}
\end{center}
\vskip .25in

\begin{abstract}
In this paper we prove a conjecture regarding the form of the
Born-Infeld Lagrangian with a $U(1)^{2n}$ gauge group  
after the elimination of the auxiliary fields.
We show that the Lagrangian
can be written as a symmetrized trace of Lorentz invariant bilinears in 
the field strength. More generally we prove a theorem regarding 
certain solutions of unilateral matrix equations of
arbitrary order. For solutions which have perturbative expansions in the  
matrix coefficients, the solution and all its positive powers 
are sums of terms which are symmetrized in all the matrix coefficients and 
of terms which are commutators. 
  
\end{abstract}
%PACS: 11.10.-q; 11.10.Jj
%\newline

\end{titlepage}

\newpage
\renewcommand{\thepage}{\arabic{page}}

\setcounter{page}{1}
\setcounter{footnote}{0}

% main text is here
\section{Duality Invariant Born-Infeld Lagrangians}
\label{BI}

In this note we prove the conjecture made in~\cite{BMZ,ABMZ}
regarding the form of
$Sp(2n, \IR)$ or $U(n,n)$
duality invariant Born-Infeld Lagrangians.
See~\cite{ABMZ} for a more extensive list of references regarding 
the duality invariance of Born-Infeld theory.
In~\cite{BMZ,ABMZ}, inspired by \cite{RT}, we exploited the fact that  the square root in 
the $U(1)$ gauge group Born-Infeld Lagrangian can be eliminated
using auxiliary fields. In the auxiliary field formulation one can
generalize the theory to a higher rank abelian gauge group $U(1)^{2n}$
such that the duality group becomes $U(n,n)$. One complication
discussed in~\cite{BMZ,ABMZ} is that one has to introduce complex gauge
fields. However  in~\cite{ABMZ} we also showed that after the
elimination of the auxiliary fields one can impose a reality
condition which preserves an  $Sp(2n, \IR)$ subgroup of the duality group.
For higher order matrices the elimination
of the auxiliary fields is more complicated since the algebraic second 
order equation for the auxiliary field 
becomes a matrix second order equation.

The Born-Infeld Lagrangian introduced in~\cite{BMZ,ABMZ} with 
auxiliary fields is given by
\[
\Lag={\rm Re}\,{\rm Tr}\,[\,\chi+
i\lambda(\chi -\frac{1}{2}  \chi  \chi^{\dagger}
+\alpha-i\beta)\,]~,
%\label{BIL}
\]
where $\alpha$ and $\beta$ are given by the following 
Lorentz invariant hermitian matrices
\[
\alpha^{ab} \equiv \frac{1}{2} F^{a} \bar{F}^{b},
~~\beta^{ab} \equiv 
\frac{1}{2}  \widetilde{F}^{a} \bar{F}^{b}.
%\label{alphadef}
\]
Here $\widetilde{F}$ is the Hodge dual of $F$ and a bar denotes
complex conjugation. The auxiliary fields $\chi$ and $\lambda$ are $n$ 
dimensional complex matrices.  
For simplicity we have set the field $S$ to the constant value $i$ since as
discussed in~\cite{ABMZ} it can be easily reintroduced. With this choice the
duality group reduces to the maximal compact subgroup 
$U(n)\times U(n)$ of $U(n,n)$.  

The equation of motion obtained by varying 
$\lambda$ gives an equation for $\chi$ 
\beqn
\chi-\frac{1}{2}\chi  \chi^{\dagger} +\alpha-i\beta=0~,
\label{EQ}
\eeqn
and after solving this equation the Lagrangian reduces to
\[
L = {\rm Re}\,{\rm Tr}\,\chi~.
\]

Let $\chi=\chi_1+i\chi_2$ where $\chi_1$ and $\chi_2$ are hermitian.
The anti-hermitian part of~\rref{EQ} implies $\chi_2 = \beta$\,, thus
$\chi^{\dagger}=\chi-2i\beta$. This can be used to eliminate $\chi$ 
from~\rref{EQ} and obtain a quadratic equation for $\chi^{\dagger}$.
Following~\cite{ABMZ}, it is convenient to define $Q=\frac{1}{2}\chi^{\dagger}$
which then satisfies
\beqn
Q=q+(p-q)Q+Q^2,
\label{Qeq}
\eeqn
where
\[
p \equiv -\frac{1}{2}(\alpha - i\beta)~, ~~ 
q \equiv -\frac{1}{2}(\alpha +i\beta) ~.
\]
The Lagrangian is then
\beqn
L =  2\, {\rm Re}\,{\rm Tr}\,Q~.
\label{Lag}
\eeqn
If the degree of the matrices is one, we can solve for $Q$ in 
the quadratic equation~\rref{Qeq} and then~\rref{Lag} 
reduces to the Born-Infeld Lagrangian.

For matrices of higher degree, equation~\rref{Qeq} can be solved
perturbatively and by analyzing the first few terms in the expansion 
we conjectured in~\cite{BMZ,ABMZ} 
that the trace of $Q$ can be obtained as follows.
First, find the perturbative solution of equation~\rref{Qeq}
assuming $p$ and $q$ commute. Then the trace of $Q$ is the trace of
the symmetrized expansion
\beqn
{\rm Tr}\,Q =\frac{1}{2}\,{\rm Tr}\left[\,
1+q-p -{\cal S}\sqrt{1-2(p+q)+(p-q)^2}\,
\right]~,
\label{CONJ}
\eeqn
where the symmetrization operator  $\cal S$ will be discussed in the
next section. 
In the appendix of~\cite{ABMZ} we have also guessed 
an explicit formula for the coefficients of the expansion of 
the trace of $Q$ 
\beqn
{\rm Tr}\,Q=
{\rm Tr}\left[\,
q+\sum_{r,s\geq 1}
\left(
\matc
r+s-2\\
r-1
\emat
\right)
\left(
\matc
r+s\\
r
\emat
\right)
{\cal S}(\,p^r q^s\,)
\,\right]~.
\label{pqLag}
\eeqn

In the next section we will 
prove that for a unilateral matrix equation of order $N$,
the perturbative solution is  
a sum of terms which are symmetrized in all the matrix coefficients and 
of  terms which are commutators. 
Since equation~\rref{Qeq} is a unilateral matrix equation the trace of
$Q$ will be symmetrized in the matrix coefficients $q$ and $p-q$.
Since this is equivalent to symmetrization in  $q$ and $p$ our
conjecture~\rref{CONJ} follows.

\section{Unilateral Matrix Equations}
\label{Mequtions}

In this section we prove a theorem regarding certain solutions of
unilateral matrix equations. These are $N^{\rm th}$ order matrix equations for the
variable $\phi$ with matrix coefficients $A_i$ which are all on one
side, e.g. on the left
\beqn
\phi=A_{0}+A_{1}\phi+A_{2}\phi^{2}+\ldots+ A_{N}\phi^{N} .
\label{eqphi}
\eeqn
The matrices are all square and of arbitrary degree. We may equally
consider the $A_i$'s as generators of an associative
algebra, and $\phi$ an element of this algebra which satisfies the
above equation. We will prove that the formal perturbative
solution of~\rref{eqphi} around zero 
is a sum of symmetrized polynomials in the $A_i$ 
and of terms which are 
commutators\footnote{If the degree of the matrices is one
the perturbative solution is convergent if $A_0$ and $A_1$ are
sufficiently small.}. The same is true for all the positive powers of
the solution.

By repeatedly inserting $\phi$ from the left hand side of~\rref{eqphi} 
into the right hand side we obtain the perturbative expansion of
$\phi$ as a sum 
\[
\phi =
\sum_{M} D_M~,
\]
where each $D_M$ is a product of the $A_i$ matrices. Any ordered 
product of these matrices will be referred to as a word. However not
every word appears in the perturbative expansion of $\phi$.  
We reserve the letter $D$ for words that do appear\footnote{This 
notation originated from an earlier version of the proof 
where the perturbative expansion of $\phi$
was calculated diagrammatically and the diagrams were denoted by $D$.
Although we will not use diagrams here, note that they are
very useful in calculating the perturbative expansion of the solution.}.

Next we obtain
the condition that a word must satisfy in order to be in the
expansion.
First note that because of~\rref{eqphi} any word $D_M$ 
can be written as the following product
\beqn
D_M=A_s D_{M_1}\ldots D_{M_s}
\label{DM}
\eeqn
for some value of $s$, where the $ D_{M_i}$'s are also words in the
expansion. Conversely, if all the $D_{M_i}$'s are words in the expansion,
$D_M$ defined in equation~\rref{DM} is also a word in the expansion.
By iterating~\rref{DM} we obtain
the following equivalent  statement: for every splitting of
$D_M$ into two words $D_M=W_1 W_2$ the second word can be written as a 
product of terms in the expansion of $\phi$
\[
D_M=W_1D_{N_1}\ldots D_{N_k}~.
\]
It is convenient to  assign to every matrix 
a dimension $d$ such that $d(\phi)=-1$. 
Using~\rref{eqphi}, the dimension of the matrix $A_i$ is given by
$d(A_i)=i-1$ and $d(D_M)=-1$.  
Then we obtain the following  intrinsic characterization of a word in
the expansion of $\phi$. It is a word $D$ such that for every splitting 
into two words $D=W_1 W_2$, where $W_2$ has at least one
letter, we have
\beqn
d(W_1) \geq 0 ~~~{\rm and}~~~ d(D)=-1~.
\label{Dcond}
\eeqn
Note that~\rref{Dcond} is a necessary and sufficient condition for a 
word to be in the expansion of $\phi$\,.

Suppose that $W$ is an arbitrary  word such that $d(W)=-1$. 
Then, as we will show, there is a 
unique cyclic permutation $D$ of $W$ such that $D$ is a term in the
expansion of $\phi$.
Let us write $W=D_{N_1}D_{N_2} \ldots D_{N_k} W_1$, where 
$D_{N_1}$ is the shortest word
starting from the first letter such that $d(D_{N_1})=-1$. 
$D_{N_i}$ is defined in the same way, except we start
from the first letter after the word $D_{N_{i-1}}$. Finally $W_1$ is
whatever is left over.
We use the notation $D_{N_i}$ since they correspond to terms in the
$\phi$ expansion.
To see this, note that 
the total dimension of a word can increase or decrease when
a letter is added on the right, but if it decreases it can only do so
by one unit. This is when the letter added is $A_0$. 
Combining this with the fact that $D_{N_i}$
is the shortest word which satisfies  $d(D_{N_i})=-1$ then implies that
if $D_{N_i}$ is a product of two
words the dimension of the first word is greater than or equal to
zero. This is just the condition~\rref{Dcond}.
Then using the fact that $d(W_1)=k-1$ one can check that 
the cyclic permutation 
of $W$ defined as  $D=W_1D_{N_1} \ldots D_{N_k}$
satisfies~\rref{Dcond}, thus it belongs to the expansion of $\phi$.
Note that all the other cyclic permutations lead to words
that are not in the expansion. Assuming the converse implies that two
distinct terms in the expansion can be related by a cyclic permutation.
But this is impossible: if we write $D=W_1W_2$, 
then $d(W_1)\geq 0$ and thus
$d(W_2)\leq -1$, so that its cyclic permutation 
$W_2W_1$ does not satisfy~\rref{Dcond}. A similar argument can be used 
to show that all different cyclic permutations of a term in the expansion of
$\phi$ lead to distinct words.

Consider the trace of the sum of all distinct words of dimension $d=-1$ and of 
order $a_i$ in $A_i$.
We can group together all words that are cyclic permutations of each other, and
replace each group by a single word with coefficient  $\sum_{i=0}^{N} a_i$. 
Using the result of the previous paragraph, we can choose 
this word to satisfy~\rref{Dcond}. Thus we have
\beqn
{\rm Tr}
\left(
\sum _{{\rm order}~\{a_i\}} D_M
\right)
~=~
\left(
  \sum_{i=0}^{N}a_i
\right)^{-1}
 {\rm Tr} 
\left(
\sum _{{\rm order}~\{a_i\}} W
\right),
\label{Distinct}
\eeqn
where the sum in the right hand side is over all distinct words of
some fixed
order $\{a_i\}$ and of dimension $d(W)=-1$.

We define 
the symmetrization operator ${\cal S}$ as a linear operator acting  
on monomials as
\beqn
{\cal S}(A_{0}^{a_0}A_{1}^{a_1} \ldots A_{N}^{a_N})
~=~
\frac{a_0!a_1! \ldots a_N!}{\left( \sum_{i=0}^{N} a_i \right)!}
\left(
\sum _{{\rm order}~\{a_i\}} W
\right)~,
\label{Sym}
\eeqn
where the sum is over distinct words of fixed order $\{a_i\}$.
Equivalently, a word can be symmetrized by averaging over all
permutations of its letters. Not all
permutations give distinct words and this accounts for  
the numerator on the right side of equation~\rref{Sym}.
The normalization of ${\cal S}$ is such that 
on commutative $A_i$'s ${\cal S}$ acts as the identity.

Combining~\rref{Distinct} and~\rref{Sym}, we
can obtain the solution for the trace of $\phi$ to all orders
\beqn
{\rm Tr}\,
\phi
~=~
\sum_{
\stackrel{\{a_i\}}
{\sum{(i-1)a_i=-1}}
}
\frac{\left( \sum_{i=0}^{N} a_i - 1 \right)!}{a_0!a_1! \ldots a_N!}
~{\rm Tr}~
{\cal S}(A_{0}^{a_0}A_{1}^{a_1} \ldots A_{N}^{a_N})~,
\label{trphi}
\eeqn 
where the sum is over all sets $\{a_i\}$ restricted to 
words of dimension $d=-1$\,.
More generally, if the $A_i$'s are considered to be the generators of
an associative algebra, we can replace the trace in~\rref{trphi} 
with the cyclic average operator which was defined in~\cite{ABMZ}.
This is true since in the proof we only used the cyclic property of
the trace which also holds for the cyclic average operator. 
Therefore, the solution $\phi$ can be written as
a sum of symmetric polynomials and terms which are commutators. 
This is the statement we set out to prove. 
Notice that our derivation implies that the 
coefficients in \rref{trphi} are all integers.

Using the same kind of arguments we used to derive equation~\rref{trphi}, 
we can also prove that the trace of positive powers of $\phi$ is given 
by
\beqn
{\rm Tr}
%\left(
\,\phi^r\,
%\right)
~=~r
\sum_{
\stackrel{\{a_i\}}
{\sum{(i-1)a_i=-r}}
}
\frac{\left( \sum_{i=0}^{N} a_i - 1 \right)!}{a_0!a_1! \ldots a_N!}
~{\rm Tr}~
{\cal S}(A_{0}^{a_0}A_{1}^{a_1} \ldots A_{N}^{a_N})~.
\label{trphir}
\eeqn 
Furthermore we can write a generating function for~\rref{trphir}
\beqn
{\rm Tr}\, \log (1-\phi)={\rm Tr}\, \log(1-\sum_{i=0}^{N} A_i)
\Big|_{d<0}~.
\label{genr}
\eeqn
On the right hand side of~\rref{genr} one must expand the logarithm
and restrict the sum to words of negative dimension. Since
$d(\phi^r)= -r$\, we can obtain~\rref{trphir} 
by extracting the dimension $d=-r$ terms from the right hand side 
of~\rref{genr}. Note that all the terms in the expansion of 
${\rm Tr}\, \log(1-\sum_{i=0}^{N} A_i)$
are automatically symmetrized.

It is possible to give a simple proof of~\rref{genr} without
going through the combinatoric arguments above, which however
give a construction of the solution and its powers themselves,
not only their trace. First note that we can
rewrite equation~\rref{eqphi} as
\[
1-\sum_{i=0}^{N} A_i
~=~
1-\phi -
\sum_{k=1}^{N} A_k (1-\phi^k)
\]
The right hand side factorizes 
\[
1-\sum_{i=0}^{N} A_i
~=~
(1-\sum_{k=1}^{N} \sum_{m=0}^{k-1} A_k\phi^m)(1-\phi)~.
\]
Under the trace we can use the fundamental property of the
logarithm, even for noncommutative objects, and  obtain
\[
{\rm Tr}\,\log(1-\sum_{i=0}^{N} A_i)
~=~
{\rm Tr}\,\log(1-\sum_{k=1}^{N} \sum_{m=0}^{k-1} A_k\phi^m)+
{\rm Tr}\,\log(1-\phi)~.
\]
Using $d(A_k)=k-1$ and $d(\phi)=-1$ we have $d(A_k\phi^m)=k-m-1$ and
we see that all the words in the argument of the first logarithm on the right hand
side have semi-positive dimension.  
Since all the words in the expansion of the second term have negative
dimension we obtain~\rref{genr}.

If the coefficient $A_N$ is unity, we have 
the following identity  for the symmetrization operator
\[
{\cal S}(A_{0}^{a_0}A_{1}^{a_1} \ldots A_{N}^{a_N})|_{A_N = 1}
=
{\cal S}(A_{0}^{a_0}A_{1}^{a_1} \ldots A_{N-1}^{a_{N-1}})~.
\]
This is obviously true up to normalization; the normalization  
can be checked in the commutative case.

The trace of the solution of~\rref{Qeq} can now be obtained
from~\rref{trphi} by 
taking $N=2$ and setting $A_2$ to unity.
The restriction on the sum of~\rref{trphi} in this case reads $a_0 -
a_2 = 1$. The sum can then be rewritten
\beqn
{\rm Tr}\,
\phi
~=~
\sum_{a_0 = 1}^{\infty}
\sum_{a_1=0}^{\infty}
\frac{\left(2a_0+a_1-2\right)!}{a_0!\,a_1!\,(a_0-1)!}
~{\rm Tr}~
{\cal S}(A_{0}^{a_0}A_{1}^{a_1})~.
\label{Trphi}
\eeqn 
Using  $\phi = Q$, $A_0 = q$, $A_1=p-q$, the combinatoric identity
\[
\left(
\matc
a+b\\
c
\emat
\right)=
\sum_{m={\rm max}(0,c-b)}^{{\rm min}(a,c)}
\left(
\matc
a\\
m
\emat
\right)
\left(
\matc
b\\
c-m
\emat
\right)
\]
and the resummation identities 
\beqarr
\sum_{r\geq1}~~~~~\sum_{a=0}^{r}~~
&=&
\sum_{a=0}^{\infty}~\sum_{r={\rm max}(a,1)}^{\infty}~,\nonumber\\
\sum_{r={\rm max}(a,1)}^{\infty}~\sum_{b=r-a+1}^{\infty}
&=&
\sum_{b={\rm max}(1,2-a)}^{\infty}~\sum_{r={\rm max}(a,1)}^{a+b-1}\nonumber
\eeqarr
one can show that~\rref{Trphi}
reduces to~\rref{pqLag}.

\section{Discussion}
After completing the first version of this paper~\cite{ABMZv1}, where we only 
proved the symmetrization theorem for the
trace of $\phi$, we learned through private communications that 
A. Schwarz was developing another method~\cite{AS} of proving the theorem (for a
slightly different, but related equation). Using his method he was able to show that the
theorem is true for arbitrary powers of the solution. 
Inspired by this, we also extended the theorem, using our method, to
positive powers of $\phi$, see~\rref{trphir}. In the process we discovered the simpler 
proof using the 
generating function~\rref{genr}.

\section*{Acknowledgments}
We would like to thank A. Schwarz for many helpful discussions.
This work was supported in part by 
the Director, Office of Science, Office of High Energy and Nuclear
Physics,~ of~ the~ U.S.~ Department~ of~ Energy~
under ~Contract 
\newline
DE-AC03-76SF00098,~ and in part by the NSF 
under grant~ PHY-95-14797. 
\newline
P.A. is supported by an INFN 
grant (concorso No. 6077/96).

\end{document}